\documentclass[runningheads,a4paper]{template/llncs}

\usepackage{amssymb}
\usepackage{graphicx}

\usepackage{url}
\urldef{\mails}\path|{bema,fcorreia,pgomes}@dei.uc.pt|
\newcommand{\keywords}[1]{\par\addvspace\baselineskip
\noindent\keywordname\enspace\ignorespaces#1}

\begin{document}

\mainmatter  

\title{Context Capture in Software Development}

\titlerunning{Context Capture in Software Development}

%
%
\author{Bruno Antunes\thanks{Supported by FCT grant SFRH/BD/43336/2008.},
Francisco Correia and Paulo Gomes}
\authorrunning{B. Antunes, F. Correia and P. Gomes}

\institute{Knowledge and Intelligent Systems Laboratory\\
Cognitive and Media Systems Group\\
Centre for Informatics and Systems of the University of Coimbra\\
Coimbra, Portugal\\
\mails\\
\url{http://www.cisuc.uc.pt/}}

%
%

\toctitle{Context Capture in Software Development}
\tocauthor{B. Antunes, P. Gomes and F. Correia}
\maketitle

\begin{abstract}
The context of a software developer is something hard to define and capture, as
it represents a complex network of elements across different dimensions that are
not limited to the work developed on an IDE. We propose the definition of a
software developer context model that takes into account all the dimensions that
characterize the work environment of the developer. We are especially focused on
what the software developer context encompasses at the project level and how it
can be captured. The experimental work done so far show that useful context
information can be extracted from project management tools. The extraction,	
analysis and availability of this context information can be used to enrich the
work environment of developers with additional knowledge to support their work.
\end{abstract}

\section{Introduction}\label{sec-introduction}

The term context has an intuitive meaning for humans, but due to this intuitive
connotation it remains vague and generalist. Furthermore, the interest in the
many roles of context comes from different fields such as literature, philosophy,
linguistics and computer science, with each field proposing its own view of
context \cite{mostefaoui_context-aware_2004}. The term context typically refers
to the set of circumstances and facts that surround the center of interest,
providing additional information and increasing understanding.

The context-aware computing concept was first introduced by Schilit and Theimer
\cite{schilit_disseminating_1994}, where they refer to context as
\emph{``location of use, the collection of nearby people and objects, as well as
the changes to those objects over time''}. In a similar way, Brown et al.
\cite{brown_context-aware_1997} define context as location, identities of the
people around the user, the time of day, season, temperature, etc. In a more
generic definition, Dey and Abowd \cite{dey_towards_2000} define context as
\emph{``any information that can be used to characterize the situation of an
entity. An entity is a person, place, or object that is considered relevant to
the interaction between a user and an application, including the user and
applications themselves''}.

In software development, the context of a developer can be viewed as a rich and
complex network of elements across different dimensions that are not limited to
the work developed on an IDE (Integrated Development Environment). Due to the
difficulty on approaching such challenge, there is not a unique notion of what
it really covers and how it can be truly exploited. With the increasing dimension
of software systems, software development projects have grown in complexity and
size, as well as in the number of functionalities and technologies involved.
During their work, software developers need to cope with a large amount of
contextual information that is typically not captured and processed in order to
enrich their work environment.

Our aim is the definition of a software developer context model that takes into
account all the dimensions that characterize the work environment of the
developer. We propose that these dimensions can be represented as a layered model
with four main layers: personal, project, organization and domain. Also, we
believe that a context model needs to be analyzed from different perspectives:
capture, modeling, representation and application. This way, each layer of the
proposed context model will be founded in a definition of what context capture,
modeling, representation and application should be for that layer.

This work is  especially focused on the project layer of the software developer
context model. We give a definition of what the context model encompasses at the
project layer and present some experimental work on the context capture
perspective.

The remaining of the paper starts with an overview of the software developer
context model we envision. In section \ref{sec-context-capture} we describe the
current work on context capture, some preliminary experimentation and the
prototype developed. An overview of related work is given in section
\ref{sec-related-work}. Finally, section \ref{sec-conclusions} concludes the
paper and point out some directions for future work.

\section{Context Model}\label{sec-context-model}

\begin{figure}
	\begin{center}
		\includegraphics[width=0.5\columnwidth,keepaspectratio=true]{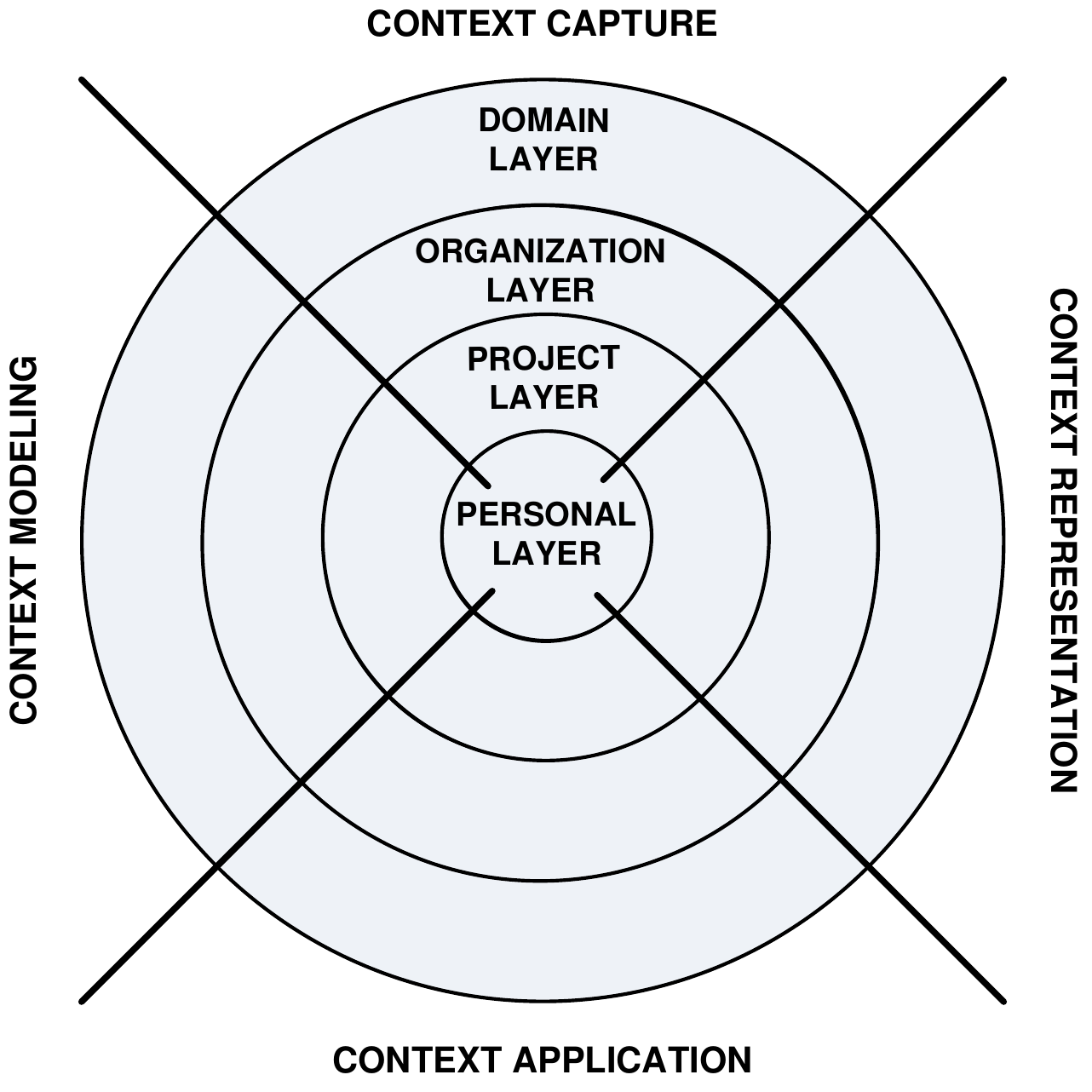}
	\end{center}
	\caption{The software developer context model layers and perspectives.}
	\label{fig-developer-context-model}
\end{figure}

The software developer context model we propose takes into account all the
dimensions that comprise the software developer work environment. This way, we
have identified four main dimensions: personal, project, organization and domain.
As shown in figure \ref{fig-developer-context-model}, these dimensions form a
layered model and will be described from four different perspectives: context
capture, context modeling, context representation and context application.

The \emph{personal layer} represents the context of the work a developer has at
hands at any point in time, which can be defined as a set of tasks. In order to
accomplish these tasks, the developer has to deal with various kinds of resources
at the same time, such as source code files, specification documents, bug
reports, etc. These resources are typically dispersed through different places
and systems, although being connected by a set of explicit and implicit relations
that exist between them. At this level the context model represents the resources
that are important for the tasks the developer is working on.

The \emph{project layer} focuses on the context of the project, or projects, in
which the developer is somehow involved. A software development project is an
aggregation of a team, a set of resources and a combination of explicit and
implicit knowledge that keeps the project running. The team is responsible for
accomplishing tasks, which end up consuming and producing resources. The
relations that exist between people and resources are the glue that makes
everything work. The project layer represents the people and resources, as well
as their relations, of the software development projects where the developer is
included.

The \emph{organization layer} takes into account the organization context to
which the developer belongs. Similarly to a project, an organization is made up
of people, resources and their relations, but in a much more complex network.
While in a project the people and resources are necessarily connected due to the
requisites of their work, in a software development organization these projects
easily become separate islands. The knowledge and competences developed in each
project may be of interest in other projects and valuable synergies can be
created when this information is available. The organization layer represents the
organizational context that surrounds a developer.

The \emph{domain layer} takes into account the knowledge domain, or domains, in
which the developer works. This layer goes beyond the project and organization
levels and includes a set of knowledge sources that stand out of these spheres.
Nowdays, a typical developer uses the Internet to search information and to keep
informed of the advances in the technologies s/he works with. These actions are
based on services and communities, such as source code repositories, development
foruns, news websites, blogs, etc. These knowledge sources cannot be detached
from the developer context and are integrated in the domain layer of our context
model. For instance, due to the dynamic nature of the software development field,
the developer must be able to gather knowledge from sources that go beyond the
limits of the organization, either to follow the technological evolution or to
search for help whenever needed.

The four context dimensions described before can be described through four
different perspectives: \emph{context capture}, which represents the sources of
context information and the way this information is gathered, in order to build
the context model; \emph{context modeling}, which represents the different
dimensions, entities and aspects of the context information (conceptual model);
\emph{context representation}, which represents the technologies and data
structures used to represent the context model (implementation); and
\emph{context application}, which represents how the context model is used and
the objectives behind its use.

\section{Context Capture}\label{sec-context-capture}

Our current work is focused on the project layer of our developer context model,
and we will discuss our work at this level from the different
perspectives we have presented before.

\begin{figure}
	\begin{center}
		\includegraphics[width=0.55\columnwidth,keepaspectratio=true]{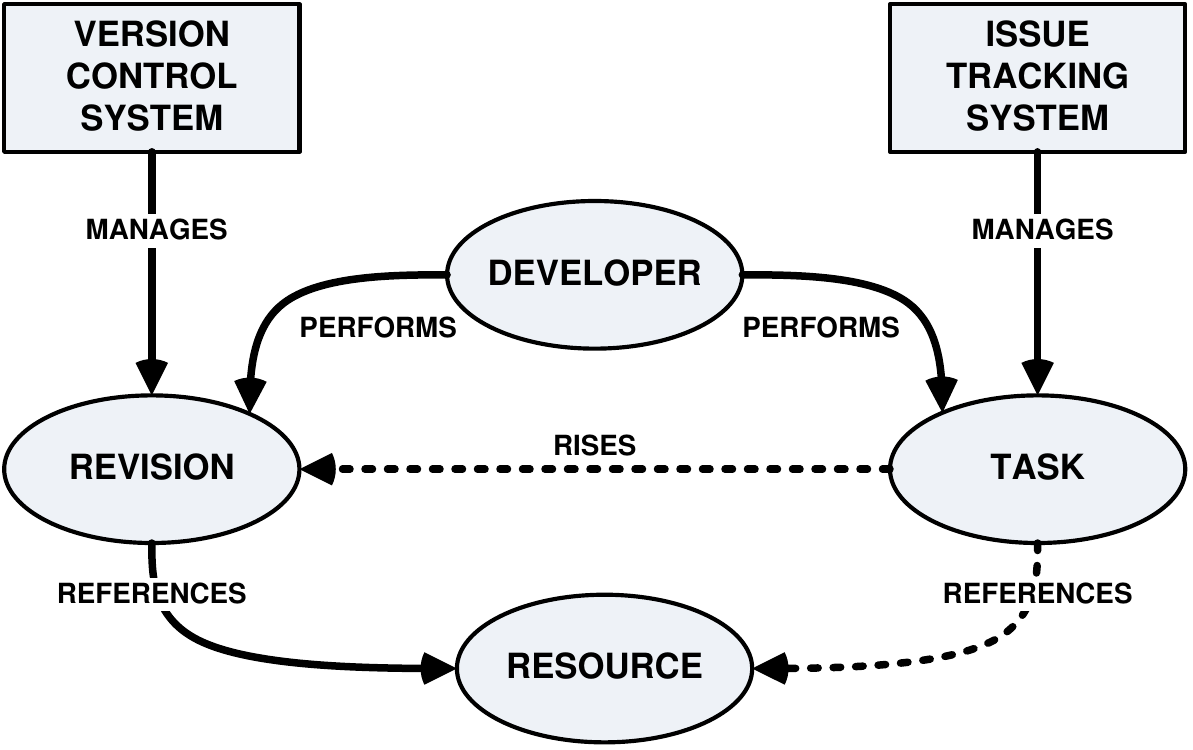}
	\end{center}
	\caption{The project layer relevant entities and their roles.}
	\label{fig-context-capture}
\end{figure}

Concerning context capture, the main sources of contextual information that feed
up the developer context model at the project level are project management tools.
These tools store a big amount of explicit and implicit information about the
resources produced during a software development project, how the people involved
relate with these resources and how the resources relate to each other. We are
focusing on two types of tools: \emph{Version Control Systems} (VCS) and
\emph{Issue Tracking Systems} (ITS). As shown in figure
\ref{fig-context-capture}, the former deals with resources and their changes, the
second deals with tasks. These systems store valuable information about how
developers, tasks and resources relate and how these relations evolve over time.
We are especially interested in revision logs and tasks. Briefly described, a
revision log tell us that a set of changes were applied by a developer to a set
of resources in the repository. A task commonly represents a problem report and
the respective fix.

\begin{figure}
	\begin{center}
		\includegraphics[width=0.5\columnwidth,keepaspectratio=true]{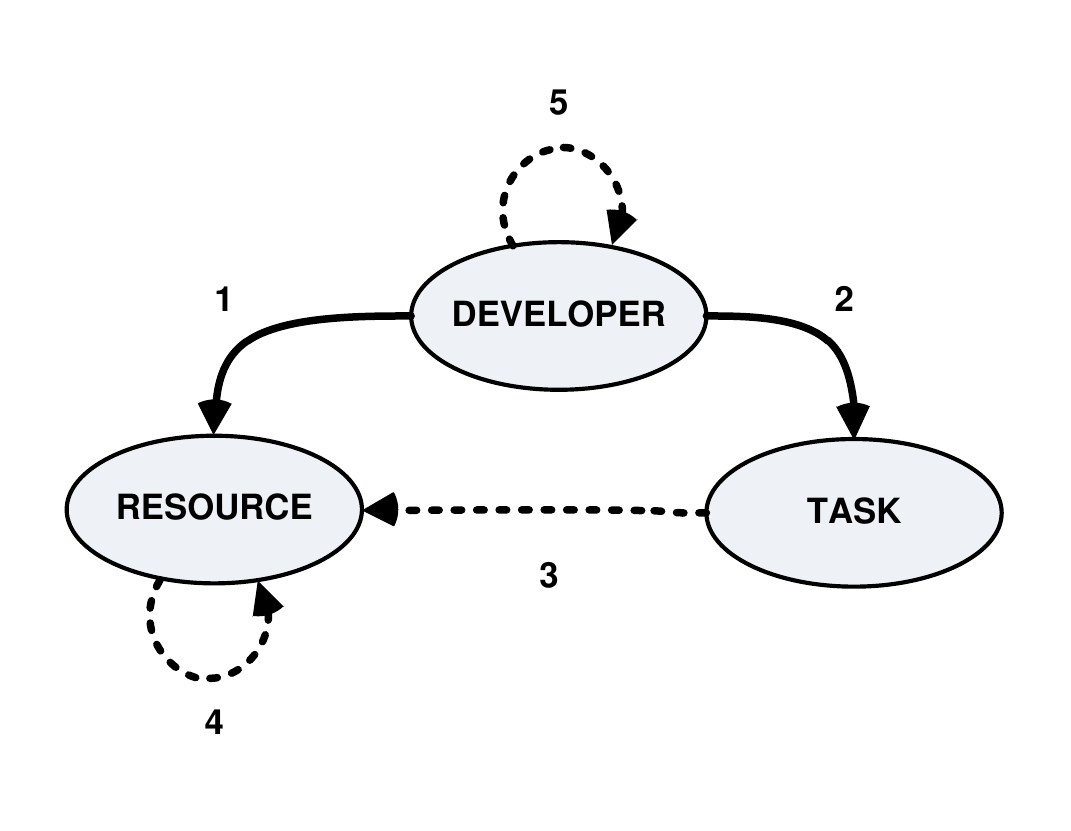}
	\end{center}
	\caption{The elements that compose the context model in the project layer.}
	\label{fig-context-modeling}
\end{figure}

The network of developers, resources, tasks and their relations will be used to
build our context model at the project level. This way, the context model of the
developer, from a project point of view, will be modeled as a set of implicit and
explicit relations, as shown in figure \ref{fig-context-modeling}. The lines with
a filled pattern represent the explicit relations and those with a dotted pattern
represent the implicit ones. The developers are explicitly related with
revisions, as they are the ones who commit the revisions, and with tasks, as
each task is assigned to a developer. The relation between tasks and resources is not
explicit, but we believe it can be identified by analyzing the information that
describe tasks and revisions. The proximity between developers can be inferred by
analyzing the resources were they share work. Also, the resources can be
implicitly related by analyzing how often they are changed together.

In order to extract relations from the information stored in project management
tools, that information is previously imported and stored locally for analysis.
The prototype developed uses a database to store both the imported data and
extracted relations. In the next phase, we intend to represent these relations
and connection elements in an ontology \cite{zuniga_ontology_2001}, which will
gradually evolve to a global developer context model ontology. We believe that
representing the context model in an ontology and formalising it using the
Semantic Web \cite{berners-lee_semantic_2001} technologies promote knowledge
sharing and reusability, since these technologies are standards accepted by the
scientific community.

The context information extracted at the project level will be used to inform the
developer about the network that links people, resources and tasks on the project
s/he works. This information can be prepared to facilitate consulting and
presented to the developer easily accessible in her/his working environment.

\subsection{Relation Extraction}

We have implemented connectors that allowed us to collect all the desired
information from the \emph{Subversion} and \emph{Bugzilla} systems, as they are
among the most popular in use. Through the collected information, we could
already perceive a set of explicit relations: which resources are created/changed
by which developers, which tasks have been assigned to which developers (see
relations number 1 and 2 in figure \ref{fig-context-modeling}). There is also a
set of implicit relations that would be valuable if disclosed.

Our approach to extract implicit relations between resources and tasks (see
relation number 3 in figure \ref{fig-context-modeling}) relies on the analysis of
the text provided by revision messages, task summaries and task comments. It is
common to find references to task identifiers in revision messages, denoting that
the revision was made in the context of a specific task. Also, task summaries and
comments commonly reference specific resources, either because a problem has been
identified in a specific class or because a error stack trace is attached to the
task summary to help developers identify the source of the problem. Taking this
into account, we have defined three algorithms to find resource/task and
task/revision relations:

\begin{itemize}
  \item \emph{Resource/Task (I)}. For each resource referenced in a revision,
  the respective name was searched in all task summaries. The search was
  performed using the file name and, in case it was a source code file, the
  full qualified name (package included) and the class name separately.
  \item \emph{Resource/Task (II)}. For each resource referenced in a revision,
  the respective name was searched in task comments. This search was performed as
  described for the previous relation.
  \item \emph{Task/Revision}. For each task, the respective identification
  number was searched in revision messages. The search was performed using common
  patterns such as \textquotedblleft \textless id\textgreater \textquotedblright,
  \textquotedblleft bug \textless id\textgreater \textquotedblright and
  \textquotedblleft \#\textless id\textgreater \textquotedblright.
\end{itemize}

The implicit relations between resources (see relation number 4 in figure
\ref{fig-context-modeling}) can be extracted by analyzing resources that are
changed together very often. Revisions are often associated with specific goals,
such as the implementation of a new feature or the correction of a bug. When
developers commit revisions, they typically change a set of resources at a time,
those that needed to be changed in order to accomplish a goal. When two resources
are changed together in various revisions, this means that these resources are
somehow related, even if they do not have any explicit relation between them,
because when one of them is modified there is a high probability that the other
also needs to be modified.

The proximity between developers (see relation number 5 in figure
\ref{fig-context-modeling}) can also be inferred by analyzing the resources they
share work. Developers can share work when they commit revisions on the same
resources or when they are assigned to tasks that are related to the same
resources. This way, if two developers often make changes, or perform tasks, on
the same resources, they are likely to be related.

\subsection{Preliminary Results}

To validate the relation extraction algorithms, these were tested against two
open-source projects from the Eclipse foundation:

\begin{itemize}
  \item \emph{gEclipse}. The gEclipse framework allows users and developers to
  access Computing Grids and Cloud Computing resources in a unified way,
  independently of the Grid middleware or Cloud Computing provider. Our analysis
  was performed over the work of 19 developers in a time window of approximately
  3 years and 3 months, which included 3279 revisions and 1508 tasks with 7765
  comments.
  \item \emph{Subversive}. The Subversive project supports the development of an
  Eclipse plug-in for Subversion integration. Our analysis was performed over the
  work of 10 developers in a time window of approximately 3 years and 2 months,
  which included 1120 revisions and 1013 tasks with 3252 comments.
\end{itemize}

By applying the relation extraction algorithms to the information related with
these two projects, we have gathered the results represented in table
\ref{tab-preliminary-results}. The table shows the number of distinct relations
extracted using each one of the algorithms in the two projects analyzed.

\begin{table}[!t]
	\renewcommand{\arraystretch}{1.3}
	\caption{Number of extracted relations.}
	\label{tab-preliminary-results}
	\centering
	\begin{tabular}{ c  |  c  |  c  |  c }
		&\bfseries Resource/Task (I)  &\bfseries Resource/Task (II) &\bfseries
		Task/Revision
		\\
		\hline
		gEclipse & 2527 & 31671 & 629 \\
		Subversive & 208 & 9076 & 773 \\
	\end{tabular} 
\end{table}

The results show that a large amount of implicit relations can be extracted from
the analysis of the information associated to tasks and revisions. These
relations complement the context model we are building by connecting tasks with
related resources. With a more detailed analysis we have identified some
problems with the algorithms creating some relations that do not correspond to
an effective correlation between the two entities analyzed. These problems are
mainly related with string matching inconsistencies and can be corrected with
minor improvements in the way expressions are searched in the text.

\subsection{Prototype}

\begin{figure}
	\begin{center}
		\includegraphics[width=0.9\columnwidth,keepaspectratio=true]{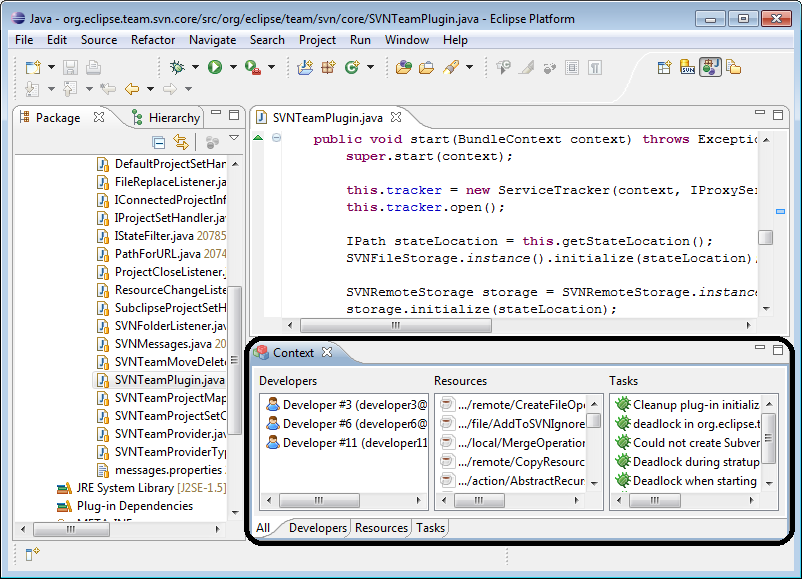}
	\end{center}
	\caption{The context plugin for Eclipse.}
	\label{fig-prototype-screenshot}
\end{figure}

We have developed a prototype, in the form of an Eclipse plug-in, to show how the
context information can be integrated into the an IDE and used to help
developers. In figure \ref{fig-prototype-screenshot} we show a screenshot of the
prototype, where we can see an Eclipse View named ``Context'' that shows context
information related to a specific resource. Each time the developer opens a
resource, this view is updated with a list of developers, resources and tasks
that are related with that resource through the relations we have described
before. This way the developer can easily gather information about what resources
are likely to be related with that resource, what other tasks affected that
resource in the past, and what other developers may be of help if extra
information is needed. The availability of this information inside the IDE, where
developers perform most of their work, increases developers awareness and reduces
their effort on finding information that would be hidden and dispersed otherwise.

\section{Related Work}\label{sec-related-work}

The EPOS (Evolving Personal to Organizational Knowledge Spaces) project,
presented by Schwarz \cite{schwarz_context_2005}, aims to build organizational
structured knowledge from information and structures owned by the elements of the
organization. The world of a knowledge worker is defined as containing
document-like objects, objects used for classification and applications. This
work environment is taken into account when modeling the user context, which
comprises information about various aspects, including currently or recently read
documents, relevant topics, relevant persons, relevant projects, etc. The context
model is formalized using RDFS \cite{brickley_rdf_2004}, and each user's context
is an up-to-date instance of this context model. The context information is
gathered and modeled through user observation and/or user feedback. The gathering
and elicitation of contextual information is done in two ways: context-polling,
by requesting a snapshot of the user's current context; and context-listening, by
subscribing the Context Service to be informed of every change in the user's
context. Being a developer a knowledge worker, much of the concepts referenced
in this work apply to the software development domain, but the specificities of
this domain demand for a context model adapted to the reality of the work
environment of a software developer.

Modularity is in the basis of the development of complex software systems and
largely used to support a programmer's tasks, but not always help the programmer
finding the desired information or delimit the point of interest for a specific
task. Based on this, Kersten and Murphy \cite{kersten_using_2006} have been
working on a model for representing tasks and their context. The task context is
derived from an interaction history that comprises a sequence of interaction
events representing operations performed on a software program's artifact. They
then use the information in a task context either to help focus the information
displayed in the IDE, or to automate the retrieval of relevant information for
completing a task. The focus of this work is the task and the knowledge elements
present in the IDE that are more relevant for the fulfillment of that task. Our
approach aims to define a context model that goes beyond the IDE and explores
the knowledge provided by the different systems that support the software
development process.

In the same line of task management and recovery, Parnin and Gorg
\cite{parnin_building_2006} propose an approach for capturing the context
relevant for a task from a programmer's interactions with an IDE, which is then
used to aid the programmer recovering the mental state associated with a task and
to facilitate the exploration of source code using recommendation systems. Their
approach is focused on analyzing the interactions of the programmer with the
source code, in order to create techniques for supporting recovery and
exploration. Again, this approach is largely restricted to the IDE and the
developer interaction with it.

With the belief that customized information retrieval facilities can be used to
support the reuse of software components, Henrich and Morgenroth
\cite{conf_dexaw_HenrichM03} propose a framework that enables the search for
potentially useful artifacts during software development. Their approach exploits
both the relationships between the artifacts and the working context of the
developer. The context information is used to refine the search for similar
artifacts, as well as to trigger the search process itself. The context
information model is represented with RDF \cite{miller_rdf_2004} statements and
covers several dimensions: the user context, the working context, and the
interaction context. While the focus here is in software reuse, our approach
focuses on the information captured during the process development.

\section{Conclusions}\label{sec-conclusions}

We have presented our approach of a software developer context model. Our context
model is based on a layered structure, taking into account four main dimensions
of the work environment of a software developer: personal, project, organization
and domain.

The current work is focused on the project layer of the software developer
context model. We have discussed this layer in more detail and presented
preliminary experimentation on the context capture perspective. The results show
that it is possible to relate tasks and revisions/resources using simple relation
extraction algorithms. These relations were then used in a plug-in for Eclipse
to unveil relevant information to the developer.

As future work we plan to improve the prototype we have developed with better
visualization, search and filtering functionality. We also want to explore the
use of ontologies to represent the developer context model. The remaining layers
of the context model will be addressed iteratively, as an extent to the work
already developed. Finally, we intend to test our approach with developers
working in with real world projects.

\bibliographystyle{template/splncs}
\bibliography{bibtex/library}

\end{document}